\documentstyle[12pt]{article}
\newlength{\extraspace}
\newlength{\extraspaces}
\newcommand{\be}{\begin{equation}
\addtolength{\abovedisplayskip}{\extraspaces}
\addtolength{\belowdisplayskip}{\extraspaces}
\addtolength{\abovedisplayshortskip}{\extraspace}
\addtolength{\belowdisplayshortskip}{\extraspace}}
\newcommand{\ee}{\end{equation}}

\newcommand{\ba}{\begin{eqnarray}
\addtolength{\abovedisplayskip}{\extraspaces}
\addtolength{\belowdisplayskip}{\extraspaces}
\addtolength{\abovedisplayshortskip}{\extraspace}
\addtolength{\belowdisplayshortskip}{\extraspace}}
\newcommand{\ea}{\end{eqnarray}}

\newcommand{\nonu}{\nonumber \\[.5mm]}
\newcommand{\A}{&\!\!\!}

\newcommand{\newsection}[1]{
\vspace{7mm} \pagebreak[3] \addtocounter{section}{1}
\setcounter{subsection}{0} \setcounter{footnote}{0}
\begin{center}
{\large {\bf \thesection. #1}}
\end{center}
\nopagebreak
\medskip
\nopagebreak \hspace{3mm}}

\begin{document}

\begin{center}
{\bf Charged Axially Symmetric Solution and Energy in Teleparallel
Theory Equivalent to General Relativity}
\end{center}
\centerline{ Gamal G.L. Nashed}

\bigskip

\centerline{{\it Mathematics Department, Faculty of Science, Ain
Shams University, Cairo, Egypt }}

\bigskip
 \centerline{ e-mail:nasshed@asunet.shams.edu.eg}

\hspace{2cm}
\\
\\
\\
\\
\\
\\
\\
\\

An exact charged solution with axial symmetry is obtained in the
teleparallel equivalent of general relativity (TEGR). The
associated metric has the structure function
$G(\xi)=1-{\xi}^2-2mA{\xi}^3-q^2A^2{\xi}^4$. The fourth order
nature of the structure function can make calculations cumbersome.
Using a coordinate transformation we get a tetrad whose metric has
the structure function  in a factorisable form
$(1-{\xi}^2)(1+r_{+}A\xi)(1+r_{-}A\xi)$ with $r_{\pm}$ as the
horizons of Reissner-Nordstr$\ddot{o}$m space-time. This new form
has the advantage that its roots are now trivial to write down.
Then, we study the  singularities of this space-time. Using
another coordinate transformation, we obtain a tetrad field. Its
associated metric yields the Reissner-Nordstr$\ddot{o}$m black
hole. In Calculating the energy content of this tetrad field using
the gravitational energy-momentum, we find that the resulting form
depends on the radial coordinate! Using the regularized expression
of the gravitational energy-momentum in the teleparallel
equivalent of general relativity we get a consistent  value for
the  energy.

\newpage
\begin{center}
\newsection{\bf Introduction}
\end{center}

The charged C-metric line element and electromagnetic potential
are given by \cite{KW} \be ds^2=\displaystyle{ 1 \over A^2(x-y)^2}
\left[ G(y) dt^2-\displaystyle{dy^2 \over G(y)} +
\displaystyle{dx^2 \over G(x)}+G(x) d\phi^2 \right], \qquad \qquad
{\cal A}=Qydt, \ee where ${\cal A}$ is the electromagnetic vector
potential, $Q$ is the charge parameter and the structure function
$G$ is defined by \be G(\xi)\stackrel{\rm def.}{=}
1-\xi^2-2mA\xi^3-q^2A^2{\xi}^4.\ee Here $m$ and $A$ are positive
parameters related to the mass and acceleration of the black hole,
such that $mA<1/\sqrt{27}$. The fact that $G$ is a fourth order
polynomial in $\xi$ means that one can not in general write down
simple expression for its roots. Since these roots play an
important role in almost every analysis of the charged C-metric,
most results have to be expressed implicitly in terms of them. Any
calculation which requires their explicit forms would naturally be
very tedious if not impossible to carry out \cite{FZ,B3,B4}.

At present, teleparallel theory seems to be popular again. There
is a trend of analyzing the basic solutions of general relativity
with teleparallel theory and comparing the results.  It is
considered as an essential part of generalized non-Riemannian
theories such as the Poincar$\acute{e}$ gauge theory \cite{Yi1}
$\sim$ \cite{Kw} or metric-affine gravity \cite{HMM}. Physics
relevant to geometry may be related to the teleparallel
description of gravity \cite{HS1,NH}. Within the framework of
metric-affine gravity, a stationary axially symmetric exact
solution of the vacuum field equations is obtained for a specific
gravitational Lagrangian by using {\it prolongation techniques}
(\cite{BH} and references therein).  Teleparallel approach is used
for positive-gravitational-energy proof \cite{Me}. A relation
between spinor Lagrangian and teleparallel theory is established
\cite{TN}.  In metric-affine generalization of teleparallelism,
Obukhov et al. \cite{OP} have shown that there is an inconsistency
in the coupling of spinors. Mielke \cite{ Me4} demonstrated the
consistency of the coupling of the Dirac fields to the TEGR.
However, Obukhov and Pereira \cite{OP} have shown that this
demonstration is not correct. They also \cite{ OP} have studied
the general teleparallel gravity model within the framework of the
metric affine gravity theory. Nester et al. \cite{NHC} have
considered the quasilocal  center-of mass (COM) in
tetrad-teleparallel gravity. They  have used the covariant
Hamiltonian formalism, in which quasilocal quantities are given by
the Hamiltonian boundary term, along with the covariant asymptotic
Hamiltonian boundary expressions.  Consideration of the COM not
only gives the most restrictive asymptotic conditions on the
variables but also gives strong constraints on the acceptable
expressions \cite{NHC}.

For a satisfactory description of the total energy of an isolated
system it is necessary that the energy-density of the
gravitational field is given in terms of first- and/or
second-order derivatives of the gravitational field variables. It
is well-known that there exists no covariant, nontrivial
expression constructed out of the metric tensor. However,
covariant expressions that contain a quadratic form of first-order
derivatives of the tetrad field are feasible. Thus it is
legitimate to conjecture that the difficulties regarding the
problem of defining the gravitational energy-momentum are related
to the geometrical description of the gravitational field rather
than are an intrinsic drawback of the theory \cite{Mj,MDTC}. M\o
ller has shown that the problem of energy-momentum complex has no
solution in the framework of gravitational field theories based on
Riemannian space-time \cite{Mo}. In a series of papers,
\cite{Mo}$\sim$\cite{Mo1} he was able to obtain a general
expression for a satisfactory energy-momentum complex in the
teleparallel space-time. Xu and Jing derived the field equation
with a cosmological term and studied the energy of the general
4-dimensional stationary axisymmetric space-time in the context of
the Hamiltonian formulation of the TEGR \cite{XJ}.

It is well known that TEGR  \cite{HS1} $\sim$ \cite{MDTC} provides
an alternative description of Einstein's general relativity. In
this theory the gravitational field is described by the tetrad
field ${e^a}_\mu$. In fact the first attempt to construct a theory
of the gravitational field in terms of  a set of four linearly
independent vector fields in the Weitzenb$\ddot{o}$ck geometry is
due to Einstein \cite{Wr,Ea}.

A well posed and mathematically consistence expression for the
gravitational energy has been developed \cite{MDTC}. It arises in
the realm of the Hamiltonian formulation of the TEGR \cite{MR} and
satisfies several crucial requirements for any acceptable
definition of gravitational energy. The gravitational
energy-momentum $P^a$ \cite{MDTC,MFC} obtained in the framework of
the TEGR has been investigated in the context of several distinct
configuration of the gravitational filed. For asymptotically flat
space-times $P^0$ yields the ADM energy \cite{ADM}. In the context
of tetrad theories of gravity, asymptotically flat space-times may
be characterized by the asymptotic boundary condition \be e_{a
\mu} \cong \eta_{a \mu} + \displaystyle{1 \over 2} h_{a
\mu}(1/r),\ee and by the condition $\partial_\mu
{e^a}_\mu=O(1/r^2)$ in the asymptotic limit $r \rightarrow
\infty$, with\\ $\eta_{a b}=(-1,+1,+1,+1)$ is the metric of
Minkowski space-time. An important property of tetrad fields that
satisfy Eq. (3) is that in the flat space-time limit one has
${e^a}_\mu(t,x,y,z)={\delta^a}_\mu$, and therefore the torsion
tensor ${T^a}_{\mu \nu}=0$. Maluf \cite{MVR} has extended the
definition $P^a$ for the gravitational energy-momentum
\cite{MDTC,MR} to any arbitrary tetrad fields, i.e., for the
tetrad fields that satisfy ${T^a}_{\mu \nu} \neq 0$ for the flat
space-time. The redefinition is the only possible consistent
extension of $P^a$, valid for the tetrad fields that do not
satisfy Eq. (3).

It is the aim of the present work to derive a charged  axially
symmetric solution in TEGR. In \S 2 we give brief review of the
TEGR of the coupled gravitational and electromagnetic fields. A
tetrad having axial symmetry with six unknown functions in $x$
$\&$ $y$ is applied to the field equations and a solution of
charged axial symmetry is obtained in \S 3. A coordinate
transformation is applied to the  tetrad obtained in \S 3, to put
the structure function in a factorisable form. The advantage of
this transformation is that it makes the roots of the original
solution be factorisable. Also in \S 3, the singularities of
tetrad (15) (see below) are studied. In \S 4, another coordinate
transformation is applied to tetrad (15) and a tetrad that its
associated metric gives the Reissner-Nordstr$\ddot{o}$m black hole
is obtained. The energy content of  tetrad (20) (also see below)
is  calculated in \S 4 using the gravitational energy-momentum
\cite{MDTC,MVR} and unsatisfactory value of energy  is obtained .
In \S 5 we use the regularized expression for the  gravitational
energy-momentum to calculate the energy.  Discussion and
conclusion of the obtained results are given in the final
section\footnote{Computer algebra system Maple 6
 is used in some calculations.}.
 \newpage
\newsection{The  TEGR for gravitation and electromagnetism}

In a space-time with absolute parallelism the parallel vector
fields ${e_a}^\mu$ define the nonsymmetric affine connection \be
{\Gamma^\lambda}_{\mu \nu} \stackrel{\rm def.}{=} {e_a}^\lambda
{e^a}_{\mu, \nu}, \ee where $e_{a \mu, \nu}=\partial_\nu e_{a
\mu}$ \footnote{space-time indices $\mu, \ \ \nu, \cdots$ and
SO(3,1) indices a, b $\cdots$ run from 0 to 3. Time and space
indices are indicated to $\mu=0, i$, and $a=(0), (i)$.} . The
curvature tensor defined by ${\Gamma^\lambda}_{\mu \nu}$ is
identically vanishing, however. The metric tensor $g_{\mu \nu}$
 is given by
 \be g_{\mu \nu}= \eta_{a b} {e^a}_\mu {e^b}_\nu. \ee

  The Lagrangian density for the gravitational field in the TEGR,
  in the presence of matter fields, is given by\footnote{Throughout this paper we use the
relativistic units$\;$ , $c=G=1$ and $\kappa=8\pi$.}
\cite{MDTC,Mjw} \be  {\cal L}_G  =  e L_G =- \displaystyle {e
\over 16\pi}  \left( \displaystyle {T^{abc}T_{abc} \over
4}+\displaystyle {T^{abc}T_{bac} \over 2}-T^aT_a
  \right)-L_m= - \displaystyle {e \over 16\pi} {\Sigma}^{abc}T_{abc}-L_m,\ee
where $e=det({e^a}_\mu)$. The tensor ${\Sigma}^{abc}$ is defined
by \be {\Sigma}^{abc} \stackrel {\rm def.}{=} \displaystyle{1
\over 4}\left(T^{abc}+T^{bac}-T^{cab}\right)+\displaystyle{1 \over
2}\left(\eta^{ac}T^b-\eta^{ab}T^c\right).\ee $T^{abc}$ and $T^a$
are the torsion tensor and the basic vector field  defined by \be
{T^a}_{\mu \nu} \stackrel {\rm def.}{=}
{e^a}_\lambda{T^\lambda}_{\mu
\nu}=\partial_\mu{e^a}_\nu-\partial_\nu{e^a}_\mu, \qquad \qquad
{T^a}_{b c} \stackrel {\rm def.}{=}  {e_b}^\mu {e_c}^\nu
{T^a}_{\mu \nu} \qquad \qquad T^a \stackrel {\rm
def.}{=}{{T^b}_b}^a.\ee The quadratic combination
$\Sigma^{abc}T^{abc}$ is proportional to the scalar curvature
$R(e)$, except for a total divergence term \cite{Mj}. $L_m$
represents the Lagrangian density for matter fields.  The
electromagnetic Lagrangian density ${\it L_{e.m.}}$ is given by
\cite{Kn} \be {\it L_{e.m.}} \stackrel {\rm def.}{=}
-\displaystyle{e \over 4} g^{\mu \rho} g^{\nu \sigma} F_{\mu \nu}
F_{\rho \sigma}, \ee where $F_{\mu \nu}$ being given
by\footnote{Heaviside-Lorentz rationalized units will be used
throughout this paper} $F_{\mu \nu} \stackrel {\rm def.}{=}
\partial_\mu A_\nu-\partial_\nu A_\mu$ and $A_\mu$ being the
electromagnetic potential.

The gravitational and electromagnetic field equations for the
system described by ${\it L_G}+{\it L_{e.m.}}$ are the following
 \ba \A \A e_{a \lambda}e_{b \mu}\partial_\nu\left(e{\Sigma}^{b \lambda \nu}\right)-e\left(
 {{\Sigma}^{b \nu}}_a T_{b \nu \mu}-\displaystyle{1 \over 4}e_{a \mu}
 T_{bcd}{\Sigma}^{bcd}\right)= \displaystyle{1 \over 2}{\kappa} eT_{a
 \mu},\nonu
  \A \A  \partial_\nu \left( e F^{\mu \nu} \right)=0, \ea
where \[ \displaystyle{ \delta L_m \over \delta e^{a \mu}} \equiv
e T_{a \mu}.\] It  is possible to prove by explicit calculations
that the left hand side of the symmetric field equations of  Eq.
(10) is exactly given by \cite{MDTC}
 \[\displaystyle{e \over 2} \left[R_{a
\mu}(e)-\displaystyle{1 \over 2}e_{a \mu}R(e) \right]. \]

\newsection{Charged Axially symmetric solution}

In this section we will assume the  parallel vector fields to have
the  form \be \left({e^a}_\mu \right) =\left(\matrix {A_1(x,y)& 0
& 0 & 0\vspace{3mm} \cr 0 & B_1(x,y)\cos\phi & 0 &
-B_2(x,y)\sin\phi \vspace{3mm} \cr 0&0&C_1(x,y)&0 \vspace{3mm} \cr
0&D_1(x,y)\sin\phi&0&D_2(x,y)\cos\phi \cr } \right), \ee where
$A_1(x,y), \ B_1(x,y), \ B_2(x,y), \ C_1(x,y), \ D_1(x,y) \ and \
D_2(x,y)$ are unknown functions. We use a non-diagonal form of the
tetrad given in Eq. (11) in spite that the metric reproduced  is
in a diagonal form. This is due to the fact that there are a
non-diagonal tetrads reproduced a diagonal metric however, a more
physics are needed to explain the obtained results
\cite{MWHL,SNH}.

Applying (11) to the field equations (10) we obtain the unknown
functions in the form \ba \A \A A_1(x,y)=\displaystyle{
\sqrt{G(y)} \over A(x-y)}, \qquad \qquad B_1(x,y)=\displaystyle{1
\over A(x-y)\sqrt{G(x)}}, \nonu
\A \A B_2(x,y)=\displaystyle{\sqrt{G(x)} \over A(x-y)}, \qquad
\qquad  C_1(x,y) =\displaystyle{1 \over  A(x-y)\sqrt{G(y)} },
\nonu
 \A \A D_1(x,y)=\displaystyle{1 \over A(x-y)\sqrt{G(x)} }, \qquad \qquad
 D_2(x,y)=\displaystyle{\sqrt{G(x)} \over A(x-y)},
   \ea where
$G(\xi)=1-{\xi}^2-2mA{\xi}^3-Q^2A^2{\xi}^4$, and the
electromagnetic potential is given by ${\cal A}= Qydt$, with
$Q=\displaystyle{q \over 2\sqrt{\pi}}$ \cite{Tn,Kn} is the charge
parameter.
  The associated metric of solution (12) has
the form (1) which is the charged C-metric. As we see in the
introduction that in general one can not easily write down simple
expression of the roots of $G$. Therefore, one must find some
coordinate transformation which makes the roots  of $G$ written
explicitly and this would in turn simplify certain analysis of the
charged C-metric. This coordinate transformation has the form
\cite{KE}

\ba \A \A x = B \left({\bar x}-c_1\right),\qquad  \qquad y =B
\left({\bar y}-c_1\right), \qquad  \phi = B_1 {\bar \phi}, \qquad
t = B_1 {\bar t},  \qquad \qquad where \nonu \A \A  B=\left(
\displaystyle { 1-{\bar A}^2{\bar Q}^2+6{\bar m}{\bar A} c_1
+6{\bar Q}^2 {\bar A}^2{c_1}^2 \over 1+(1-{\bar Q}^2{\bar
A}^2){c_1}^2+4{\bar m} {\bar A} {c_1}^3+3{\bar Q}^2{\bar
A}^2{c_1}^4 } \right)^{1/2} \nonu
 \A \A  B_1= \sqrt{1-{\bar A}^2{\bar Q}^2+6{\bar m}{\bar A} c_1
+6{\bar Q}^2 {\bar A}^2{c_1}^2} \sqrt{1+(1-{\bar Q}^2{\bar
A}^2){c_1}^2+4{\bar m} {\bar A} {c_1}^3+3{\bar Q}^2{\bar
A}^2{c_1}^4 }, \ea with ${\bar x}, {\bar y}, {\bar \phi}, {\bar
t}$ are the new coordinate and \ba \A \A Q=\displaystyle {{\bar Q}
\over 1-{\bar A}^2{\bar Q}^2+6{\bar m}{\bar A} c_1 +6{\bar Q}^2
{\bar A}^2{c_1}^2}, \qquad \qquad  m=\displaystyle {{\bar
m}+2{\bar Q}^2 {\bar A}{c_1} \over (1-{\bar A}^2{\bar Q}^2+6{\bar
m}{\bar A}c_1+ 6{\bar Q}^2 {\bar A}^2{c_1}^2)^{3/2}}, \nonu
 \nonu
\A \A  A=\sqrt{1+(1-{\bar Q}^2{\bar A}^2){c_1}^2+ 4{\bar m} {\bar
A} {c_1}^3+3{\bar Q}^2{\bar A}^2{c_1}^4 }{\bar A},\nonu
\nonu %
 \A \A 4{\bar Q}^2{\bar A}^2{c_1}^3+6{\bar m} {\bar A}
{c_1}^2+2(1-{\bar Q}^2{\bar A}^2) c_1=2{\bar m} {\bar A}. \ea
Applying the coordinate transformation (13) to the tetrad (11)
with solution (12) we obtain \be  \left({e^a}_\mu \right)
=\left(\matrix {\displaystyle{H({\bar y}) \over {\bar A} ({\bar
x}-{\bar y})}
 & 0  & 0 & 0 \vspace{3mm} \cr
0 & \displaystyle{\cos {\bar \phi}^{\ast} \over {\bar A}({\bar
x}-{\bar y})H({\bar x})}&
   0& -\displaystyle{H({\bar x})\sin {\bar \phi}^{\ast} \over
   {\bar A}({\bar x}-{\bar y})}
\vspace{3mm} \cr 0&0&\displaystyle{1 \over {\bar A} ({\bar x}
-{\bar y})H({\bar y})} &0 \vspace{3mm} \cr
 0& \displaystyle{\sin {\bar \phi}^{\ast} \over {\bar A}({\bar x}-{\bar y})H({\bar x})}
 &0& \displaystyle{\displaystyle{\cos {\bar \phi}^{\ast}H({\bar x}) \over
{\bar A}({\bar x}-{\bar y})} }\cr } \right), \ee  where
\[{\bar \phi}^\ast=B_1{\bar \phi}
\] with $B_1$  given in Eq. (13) and
\[ H(\xi)=\sqrt{1-\xi^2+{\bar Q}^2{\bar A}^2\xi^2+2{\bar m}{\bar A} \xi-
2{\bar m}{\bar A} \xi^3 -{\bar Q}^2{\bar A}^2\xi^4}.\] The
associated metric of the tetrad field given by Eq. (15) is given
by \ba ds^2 \A =\A \displaystyle{ 1 \over A^2(x-y)^2} \left[
G_1(y) dt^2-\displaystyle{dy^2 \over G_1(y)} + \displaystyle{dx^2
\over G_1(x)}+G_1(x) d\phi^2\right],   \ where \ G_1(\xi) \ is \
defined \ by \nonu
\A \A  G_1(\xi) \stackrel{\rm def.}{=}
(1-{\xi}^2)(1+r_{+}A\xi)(1+r_{-}A\xi)=H^2(\xi), \quad with \quad
r_{\mp}={\bar m}\mp\sqrt {{\bar m}^2-{\bar Q}^2},\ea  and $0\leq
r_{-}A \leq r_{+} A<1$ \cite{KE} and the electromagnetic potential
has the form
\[ {\cal A}={\bar Q}({\bar x}-c_1)dt.\] It is clear from (16) that one
can gets the roots easily which has the form \be
\xi_{1,2}=-\displaystyle{1 \over r_{\mp} A}, \qquad
 \xi_{3,4}=\mp 1 \qquad which
\quad obey \quad \xi_1\leq \xi_2<\xi_3<\xi4.\ee  Now we are going
to study the singularities of tetrad (15).

In teleparallel theories of gravity we mean by singularity of
space-time \cite{Kn} the singularity of the scalar concomitants
of the curvature and torsion tensors.

Using the definitions  of the  Riemann Christoffel, Ricci tensors,
Ricci scalar, torsion tensor and  basic vector Eq. (8), \cite{Ncs}
we obtain for  solution (15)
 \ba
 R^{\mu \nu \lambda \sigma}R_{\mu \nu \lambda
\sigma} \A = \A  F_1({\bar x},{\bar y}), \qquad \qquad R^{\mu
\nu}R_{\mu \nu}= F_2({\bar x},{\bar y}), \qquad  \qquad
R=F_3({\bar x},{\bar y}) ,\nonu
T^{a b c}T_{a b c} \A=\A \displaystyle{ F_4({\bar x},{\bar y})
\over (1-{\bar x}^2)(1-{\bar y}^2)(1+2{\bar x}{\bar m}{\bar
A}+{\bar Q}^2{\bar x}^2{\bar A}^2)(1+2{\bar y}{\bar m}{\bar
A}+{\bar Q}^2{\bar y}^2{\bar A}^2)}, \nonu
T^a T_a \A=\A  \displaystyle{F_5({\bar x},{\bar y}) \over (1-{\bar
x}^2)(1-{\bar y}^2)(1+2{\bar x}{\bar m}{\bar A}+{\bar Q}^2{\bar
x}^2{\bar A}^2)(1+2{\bar y}{\bar m}{\bar A}+{\bar Q}^2{\bar
y}^2{\bar A}^2)},
 \ea
 where $F_i, \ i=1 \cdots 5$ are too lengthy functions of ${\bar x}$
 and ${\bar y}$.
 It  is clear from (18) that the scalars of torsion and  basic vector
  have the same singularities as the dominator of both are the same. Let us discuss
 these singularities.\\ 1) When ${\bar x}={\bar y}=\xi_3$ then all the scalars of (18)
 have a singularities which is called {\it asymptotic infinity} \cite{KE}
  .\\ 2) When ${\bar y}=\xi_2$, there is a singularity
 which is called {\it black hole event horizon} \cite{KE}.\\3)
 When ${\bar y}=\xi_3$ there is also a singularity which is {\it acceleration
 horizon}.\\ 4) When ${\bar x}=\xi_4$ there is a singularity which makes
 {\it symmetry axis between event and acceleration horizons}.\\ 5) When
 ${\bar x}=\xi_3$ there is a singularity which makes {\it a symmetry axis
 joining between event horizon with asymptotic horizon}.\\ 6) When
 ${\bar x}=\xi_3$ and ${\bar y}=\xi_4$ there will be  {\it a conical singularity} \cite{KE}.
\newsection{Energy content }
To write the tetrad field given in Eq. (15) into a spherical polar
coordinate, we will use the following coordinate transformation
\cite{KE} \be {\bar x}=\cos \theta, \qquad \qquad {\bar
y}=-\displaystyle{1 \over {\bar A} r}, \qquad \qquad {\bar
\phi}^{\ast}={\bar \phi}^{\ast}, \qquad \qquad {\bar t}={\bar
A}t_1, \ee where $r, \theta, t_1$ are the new coordinate. Applying
transformation (19) to the tetrad field (15) we get \be
\left({e^a}_\mu \right) =\left(\matrix
{-\displaystyle{\sqrt{r^2-2{\bar m}r+{\bar Q}^2} \ H_1 \over r\
G_2}&0 &0 &0\vspace{3mm} \cr 0&0 &\displaystyle{r \cos{\bar
\phi}^\ast \over F\ G_2} &\displaystyle{r \ F \sin\theta \sin{\bar
\phi}^\ast \over G_2} \vspace{3mm} \cr 0&\displaystyle{r \over
\sqrt{r^2-2{\bar m}r+{\bar Q}^2} G_2 \ H_1} &0&0 \vspace{3mm} \cr
0& 0 &\displaystyle{r \sin{\bar \phi}^\ast \over F\ G_2}
&-\displaystyle{r\ F \sin\theta \cos{\bar \phi}^\ast \over G_2}
\cr } \right), \ee where
\[F=\sqrt{1+2{\bar m}{\bar
A}\cos \theta +{\bar Q}^2{\bar A}^2 \cos^2 \theta}, \qquad
G_2=({\bar A}r \cos \theta+1), \qquad H_1=\sqrt{{\bar
A}^2r^2-1}.\] Taking the $limit_{{\bar A} \rightarrow 0}$ in (20),
the associate metric will have the Reissner-Nordstr$\ddot{o}$m
space-time. Now we are going to calculate the energy content of
Eq. (20). Before this let us give a brief review of the derivation
of the gravitational energy-momentum.

Multiplication of the symmetric part of Eq. (10) by the
appropriate inverse tetrad fields yields it to have the form
\cite{MDTC,MVR}  \be
\partial_\nu \left(-e {\Sigma}^{a \lambda \nu } \right)=-\displaystyle{e
e^{a \mu} \over 4} \left(4{\Sigma}^{b \lambda \nu }T_{b \nu \mu }-
{\delta^\lambda}_\mu {\Sigma}^{bdc}T_{bcd} \right)-4\pi {e^a}_\mu
T^{\lambda \mu}.\ee By restricting the space-time index $\lambda$
to assume only spatial values then Eq. (21) takes the form
\cite{MDTC} \be \partial_0 \left(e {\Sigma}^{a 0 j}
\right)+\partial_k\left(e {\Sigma}^{a k j}
\right)=-\displaystyle{e e^{a \mu} \over 4}\left(4{\Sigma}^{b c
j}T_{b c \mu }- {\delta^j}_\mu {\Sigma}^{bcd}T_{bcd} \right)-4\pi
e {e^a}_\mu T^{j \mu}.\ee Note that the last two indices of
${\Sigma}^{abc}$ and $T^{abc}$ are anti-symmetric. Taking the
divergence of Eq. (22) with respect to j yields

\be -\partial_0 \partial_j \left(-\displaystyle {1  \over 4 \pi} e
{\Sigma}^{a 0 j} \right)=-\displaystyle{1 \over 16 \pi}
\partial_j\left[e e^{a \mu}\left(4{\Sigma}^{b c j}T_{b c \mu }-
{\delta^j}_\mu {\Sigma}^{bcd}T_{bcd} \right)-16 \pi( e {e^a}_\mu
T^{j \mu})\right].\ee

In the Hamiltonian formulation of the TEGR \cite{BN,MR} the
momentum canonically conjugated to the tetrad components $e_{aj}$
is given by \[ \Pi^{a j}=-\displaystyle {1  \over 4 \pi} e
{\Sigma}^{a 0 j},\] and that the gravitational energy-momentum
$P^a$ contained within a volume $V$ of the three-dimensional
spacelike hypersurface is defined by \cite{MDTC} \be P^a=-\int_{V}
d^3 x
\partial_j \Pi^{a j}.\ee If no condition is imposed on the tetrad
field, $P^a$ transforms as a vector under the global $SO(3,1)$
group. It describes the gravitational energy-momentum with respect
to observers adapted to ${e^a}_\mu$. This observers are
characterized by the velocity field $u^\mu={e_{(0)}}^\mu$ and by
the acceleration $f^\mu$
\[ f^\mu=\displaystyle{Du^\mu \over ds}=\displaystyle{D{e_{(0)}}^\mu \over
ds}=u^a \nabla_a {e_{(0)}}^\mu.\] Let us assume that the
space-time is asymptotically flat. The total
 gravitational energy-momentum is given by \be P^a=-\oint_{S \rightarrow \infty}
 dS_k \Pi^{a k}.\ee The field quantities are evaluated on a
 surface $S$ in the limit $r \rightarrow \infty.$

 Now we are going to apply Eq. (25) to the tetrad field (20) to calculate
 the energy content. We perform the calculations in the Cartesian coordinate.
  Eqs. (24) and (25)
 assumed that the reference space is determined by a set of tetrad
 fields ${e^a}_\mu$ for the flat space-time such that the
 condition ${T^a}_{\mu \nu}=0$ is satisfied.
 It is clear from (22) that the only components which contributes to the energy is
 $\Sigma^{0 0 \alpha}$. Thus substituting from solution (20) into
 (7), we obtain the following non-vanishing value
 \ba
{\Pi}^{0 a} \A \cong  \A {x^a \over \kappa
r^3(x^2+y^2)\left(1-\displaystyle{2{\bar m} \over
 r}+\displaystyle{q^2 \over r^2}\right)}\left(r^3+\left\{r-6{\bar m}
 +\displaystyle{3{\bar Q}^2 \over
r}\right \}(x^2+y^2) \right), \qquad a=1,2, \nonu
 {\Pi}^{0 3} \A \cong \A {z \over \kappa r^3\left(1-\displaystyle{2{\bar m} \over
 r}+\displaystyle{{\bar Q}^2 \over r^2}\right)
 }\left(r-6{\bar m}+\displaystyle{3{\bar Q}^2 \over r}
\right).
 \ea
 Substituting from (26) into
(25) we get the form of energy contained within a sphere of radius
$R$ given by\be P^{(0)}=E(R)=-R \left({1-\displaystyle{2{\bar m}
\over  R}+\displaystyle{{\bar Q}^2 \over
R^2}}\right)^{-3/2}\left(1-\displaystyle{4{\bar m} \over R}
 +\displaystyle{2{\bar Q}^2 \over R^2}\right)
  \cong -\left(R-{\bar m}+\displaystyle{{\bar Q}^2 \over 2R}\right). \ee
  \newsection{Regularized expression for the gravitational energy-momentum
  and localization of energy}

An important property of the tetrad fields that satisfy the
condition of Eq. (3) is that in the flat space-time limit
${e^a}_\mu(t,x,y,z)={\delta^a}_\mu$, and therefore the torsion
${T^\lambda}_{\mu \nu}=0$.  Hence for the flat space-time it is
normally to consider a set of tetrad fields such that
${T^\lambda}_{\mu \nu}=0$ {\it in any coordinate system}. However,
in general an arbitrary set of tetrad fields that yields the
metric tensor for the asymptotically flat space-time does not
satisfy the asymptotic condition given by (3). Moreover for such
tetrad fields the torsion ${T^\lambda}_{\mu \nu} \neq 0$ for the
flat space-time \cite{MVR}. It might be argued, therefore, that
the expression for the gravitational energy-momentum (24) is
restricted to particular class of tetrad fields, namely, to the
class of frames such that ${T^\lambda}_{\mu \nu}=0$ if ${E^a}_\mu$
represents the flat space-time tetrad field \cite{MVR}. To explain
this, let us calculate the flat space-time of the tetrad field of
Eq. (20) which is given by \be \left({E^a}_\mu \right)
=\left(\matrix {1&0 &0 &0 \vspace{3mm} \cr 0&0 & -r \cos{\bar
\phi}^\ast&-r\sin\theta \sin{\bar \phi}^\ast \vspace{3mm} \cr 0&
1&0&0 \vspace{3mm} \cr 0&0&-r\sin{\bar \phi}^\ast& r\sin \theta
\cos{\bar \phi}^\ast \cr } \right). \ee Expression (28) yields the
following non-vanishing torsion components: \ba  \A \A T_{(1) 2
1}=-\cos{\bar \phi}^\ast, \qquad T_{(1) 3 1}=-\sin\theta \cos{\bar
\phi}^\ast, \qquad T_{(1) 2 3}=r \sin{\bar
\phi}^\ast(\cos\theta+1), \nonu
\A \A T_{(3) 12}=\sin{\bar \phi}^\ast, \qquad
T_{(3)13}=-\sin\theta \sin{\bar \phi}^\ast, \qquad
T_{(3)23}=-r\cos{\bar \phi}^\ast(\cos(\theta)+1).\ea The tetrad
field (28) when written in the Cartesian coordinate will have the
form \be \left({E^a}_\mu(t,x,y,z) \right) =\left(\matrix {1&0 &0
&0 \vspace{3mm} \cr 0&\displaystyle{y^2r-x^2z \over r(x^2+y^2)} &
-\displaystyle{yx(z+r) \over r(x^2+y^2)} &\displaystyle{x \over r}
\vspace{3mm} \cr 0& \displaystyle{x \over r} &\displaystyle{y
\over r} &\displaystyle{z \over r} \vspace{3mm} \cr
0&-\displaystyle{yx(z+r) \over r(x^2+y^2)}&\displaystyle{x^2r-y^2z
\over r(x^2+y^2)}& \displaystyle{y \over r} \cr } \right). \ee In
view of the geometric structure of (30), we see that, Eq. (20)
does not display the asymptotic  behavior required by Eq. (3).
Moreover, in general the tetrad field (30) is adapted to
accelerated observers \cite{MDTC,MR,MVR}. To explain this, let us
consider a boost in the x-direction of Eq. (30).  We find \be
\left({E^a}_\mu(t,x,y,z) \right) =\left(\matrix {\gamma&-v \gamma
&0 &0 \vspace{3mm} \cr -v \gamma & \gamma\displaystyle{y^2r-x^2z
\over r(x^2+y^2)} & -\displaystyle{yx(z+r) \over r(x^2+y^2)}
&\displaystyle{x \over r} \vspace{3mm} \cr 0& \displaystyle{x
\over r} &\displaystyle{y \over r} &\displaystyle{z \over r}
\vspace{3mm} \cr 0&-\displaystyle{yx(z+r) \over
r(x^2+y^2)}&\displaystyle{x^2r-y^2z \over r(x^2+y^2)}&
\displaystyle{y \over r}  \cr } \right), \ee where $v$ is the
speed of the observer and $\gamma=\sqrt{1-v^2}$. It can be shown
that along an observer's trajectory whose velocity is determined
by \be {\it u}^u=E_{(0)}^\mu=(\gamma, -v\gamma,0,0), \quad the
\quad quantities \quad {\phi_{(j)}}^{(k)}=u^i\left({E^{(k)}}_m
\partial_i {E_{(j)}}^m\right),\ee constructed out from (31) are
non vanishing. This fact indicates that along the observer's path
the spatial axis ${E_{(a)}}^\mu$ rotate \cite{MR,MVR}. In spite of
the above problems discussed for the tetrad field of Eq. (20) it
yields a satisfactory value for the total gravitational
energy-momentum, as we will discussed.

 In Eqs. (24) and (25) it is implicitly assumed that the reference space is determined
 by a set of tetrad fields ${e^a}_\mu$ for flat space-time such
 that the condition ${T^a}_{\mu \nu}=0$ is satisfied. However, in
 general there exist flat space-time tetrad fields for which ${T^a}_{\mu \nu} \neq
 0$. In this case Eq. (24) may be generalized \cite{MR,MVR} by
 adding a suitable reference space subtraction term, exactly like
 in the Brown-York formalism \cite{BHS,YB}.

 We will denote ${T^a}_{\mu \nu}(E)=\partial_\mu {E^a}_\nu-\partial_\nu
 {E^a}_\mu$ and $\Pi^{a j}(E)$ as the expression of $\Pi^{a j}$
 constructed out of the flat tetrad ${E^a}_\mu$. {\it The
 regularized form of the gravitational energy-momentum $P^a$ is
 defined by} \cite{MR,MVR}
 \be P^a=-\int_{V} d^3x \partial_k \left[ \Pi^{a k}(e)-\Pi^{a k}(E)
 \right].\ee This condition guarantees that the energy-momentum of
 the flat space-time always vanishes. The reference space-time is
 determined by tetrad fields ${E^a}_\mu$, obtained from
 ${e^a}_\mu$ by requiring the vanishing of the physical parameters
 like mass, angular momentum, etc. Assuming that the space-time is
 asymptotically flat then Eq. (33) can have the form \cite{MR,MVR}

\be P^a=-\oint_{S\rightarrow \infty} dS_k \left[ \Pi^{a
k}(e)-\Pi^{a k}(E) \right],\ee where the surface $S$ is
established at spacelike infinity. Eq. (34) transforms as a vector
under the global SO(3,1) group. Now we are in a position to proof
that the tetrad field (20) yields a satisfactory value for the
total gravitational energy-momentum.

We will integrate Eq. (34) over a surface of constant radius
$x^1=r$ and require $r\rightarrow \infty$. Therefore, the index
$k$ in (34) takes the value $k=1$. We need to calculate the
quantity \[\Sigma^{(0) 01}={e^{(0)}}_0\Sigma^{0
01}=\displaystyle{1 \over 2}{e^{(0)}}_0(T^{001}-g^{00}T^{1}).\]
Evaluate  the above equation we find \be
-\Pi^{(0)1}(e)=\displaystyle{1 \over 4\pi}e\Sigma^{(0)
01}=-\displaystyle{1 \over 4\pi}r\sin(\theta)\sqrt{1-{2{\bar m}
\over r}+{{\bar Q}^2 \over r^2}},\ee and the expression of
$\Pi^{(0)1}(E)$ is obtained by just making ${\bar m}=0$ and ${\bar
Q}=0$ in Eq.(35), it is given by \be\Pi^{(0)1}(E)=\displaystyle{1
\over 4\pi}r\sin(\theta).\ee Thus the gravitational energy
contained within a sphere of radius $R$ is given by  \be P^0 \cong
\int_{r\rightarrow R} d\theta d\phi \displaystyle{1 \over
4\pi}\sin(\theta)\left\{-r(1-{{\bar m} \over r}+{{\bar Q}^2 \over
2r^2})+r\right\}={\bar m}-{{\bar Q}^2 \over 2R},\ee which is the
expected result.
\newsection{Main results and Discussion}
The main results of this paper are the following:\vspace{0.4cm}\\
$\bullet$  Simple expression of the roots of the structure
function
has been obtained in Eq. (16).\vspace{0.3cm}\\
$\bullet$  The singularities of the tetrad field of Eq. (15)  are
shown to be  related to the roots of the structure function.\vspace{0.3cm}\\
$\bullet$ The tetrad field given in Eq. (15) with $(t, x,y,{\bar
\phi}^\ast)$ has been  transformed to spherical polar coordinate
with $(t,r,\theta,{\bar \phi}^\ast)$.\vspace{0.3cm}\\
 $\bullet$ Setting the physical parameters equal to zero in the tetrad
 field given in Eq. (20), i.e. \\ ${\bar m}=0 \ and \ {\bar Q}=0$, we have
 obtained a non Minkowskian space-time.\vspace{0.3cm}\\
$\bullet$ It is well know that calculations of global energies and
momenta in TEGR are much easier than in GR. Therefore, we have
used the regularized expression of the gravitational
energy-momentum given in Eq. (34) to calculate the mass-energy
given by Eq. (37).

\vspace{2cm} \centerline{\large {\bf Acknowledgment}} The author
would like to  thank the Referee for careful reading and for
putting the paper in  a more readable form. Aslo author
 would like to thank  Professor J.G. Pereira  Universidade Estadual
 Paulista, Brazil and Professor J.W. Maluf Instituto de F${\acute i}$sica,Universidade
 de Bras${\acute i}$lia Brazil.

\end{document}